\documentclass[%
aps,%
prb,%
reprint,%
amsmath,amssymb,%
floatfix,%
nofootinbib,%
superscriptaddress%
]{revtex4-2}

\usepackage[utf8]{inputenc}
\usepackage{amsmath,amsthm}
\usepackage{graphicx}
\usepackage[colorlinks = true,
            linkcolor = blue,
            urlcolor  = blue,
            citecolor = blue,
            anchorcolor = blue]{hyperref}
\usepackage{braket}
\usepackage{xcolor}
\usepackage{enumitem}
\newcommand{\rme}{\mathrm{e}}

\newcommand{\I}{\mathbf{I}}

\newcommand{\bea}{\begin{eqnarray}}
\newcommand{\eea}{\end{eqnarray}}

\global\long\def\ga{\gamma} \global\long\def\de{\delta}

\global\long\def\ell#1{\theta_{#1}}

\global\long\def\la{\lambda} \global\long\def\ka{\kappa}
\global\long\def\si{\sigma}
\global\long\def\vfi{\varphi}

\global\long\def\al{\alpha}
\global\long\def\be{\beta}
\global\long\def\ga{\gamma} \global\long\def\de{\delta}

\global\long\def\no{\nonumber}
\renewcommand{\emph}[1]{\textbf{\textit{#1}}}

\makeatletter
\newcommand{\tagsleft}{\tagsleft@true}
\newcommand{\tagsright}{\tagsleft@false}
\makeatother

\newcommand{\ba}{\begin{array}}
\newcommand{\ea}{\end{array}}

\newcommand{\bel}[1]{\begin{equation}\label{#1}}
\usepackage{dsfont}

\theoremstyle{thm@}

\theoremstyle{remark}

\def\beq{\begin{equation}}
\def\enq{\end{equation}}

\newcommand{\vb}{\mathbf{v}}
\global\long\def\dG{\mathbb{G}}
\newcommand{\exval}[1]{\mbox{$\langle \, {#1}\, \rangle$}}

\newcommand{\rmd}{\mathrm{d}}

\newcommand{\R}{{\mathbb R}}

\begin{document}
\title{Robust Golden Ratio Universality as a consequence of a circulant symmetry}

\author{Luka Paljk}
\email{luka.paljk@fmf.uni-lj.si}
\affiliation{Faculty of Mathematics and Physics, University of Ljubljana, Jadranska 19, SI-1000 Ljubljana, Slovenia}
\author{Johannes Schmidt}
\affiliation{Bonacci GmbH, Robert-Koch-Str. 8, 50937 Cologne, Germany}
\author{Vladislav Popkov}
\affiliation{Faculty of Mathematics and Physics, University of Ljubljana, Jadranska 19, SI-1000 Ljubljana, Slovenia}
 \affiliation{Department of Physics,
  University of Wuppertal, Gaussstra\ss e 20, 42119 Wuppertal,
  Germany}

\begin{abstract}
We find a robust scenario for the emergence of the Golden Ratio (GR) dynamical universality class, characterized by the dynamical exponent $z\equiv\varphi=\frac{1}{2}(1+\sqrt{5})$. Previously, its emergence required fine-tuning of system parameters. Here, the GR universality class appears generically in a wide class of Markov processes as a consequence of a global, circulant symmetry of the flux Jacobian. This symmetry leads to GR universality in systems where the total number of conservation laws is odd and larger than three. We introduce a multilane exclusion process with Bernoulli stationary measure as a minimal, parameter-free model, which features GR universality for all conserved modes, most of which are doubly degenerate. Our predictions, based on mode-coupling theory generalized to treat degenerate conserved modes, are corroborated by large-scale Monte Carlo simulations.
\end{abstract}

\maketitle

\emph{Introduction.---}
A remarkable universality class describing the relaxation of long-lived modes in driven systems, characterized by the Golden Ratio (GR) dynamical exponent $z\equiv \varphi=(1+\sqrt{5})/2$,  was predicted in Ref.~\cite{2015SpohnStoltz} based on mode-coupling theory (MCT). 
Within the framework of nonlinear fluctuating hydrodynamics (NLFH) \cite{Spoohn14}, MCT predicts that
systems with multiple conservation laws exhibit a discrete family of dynamical
exponents given by ratios of consecutive Fibonacci numbers~\cite{2015Fibonacci},
$z_n = F_{n+1}/F_n$, where $n \geq 2$ and $F_n$ is the $n$-th Fibonacci number $(1,1,2,3,5,8,13,21,\ldots)$.
As $n\to\infty$, these ratios converge to the Golden Ratio, $z\to\varphi=(1+\sqrt{5})/2$ \cite{livio2008golden}.
Only two members of the hierarchy are robust: the diffusive class \cite{EW},
$z = 2 = F_3/F_2$, and the KPZ class \cite{KPZ}, $z = 3/2 = F_4/F_3$. The remaining
universality classes require fine tuning, as they demand several conserved
modes together with specific couplings among them \cite{2015Fibonacci}. The GR
universality class, for instance, arises only when two hydrodynamic modes
couple to each other but neither couples to itself. More generally, in a
system with an arbitrary number of conserved modes, GR universality  for all modes is
realized  provided that each mode is coupled to some other mode (or modes) and
\textit{not} to itself.
Although the conditions for the GR universality class are straightforward to formulate, they appear to require fine-tuning and are therefore difficult to realize naturally. In Ref.~\cite{2015Fibonacci} these conditions required the fine-tuning of \textit{six} independent interaction parameters characterizing the hopping rates in a system with only three conserved modes.
Furthermore, a detailed analysis of the minimal scenario of two coupled GR modes yielded mostly no-go statements, showing that the GR universality class is highly non-generic \cite{2024Golden}.

In this Letter, we show that the GR universality class can be realized robustly through a \textit{global circulant symmetry} of the system, rather than by fine-tuning microscopic parameters.
Once this symmetry is satisfied, the GR scenario is guaranteed, independently of the microscopic details of the model in systems with an odd number of conservation laws $M \geq 5$.  
A natural realization of the circulant symmetry is that of Markov processes with a multilane geometry in which the particle number is separately conserved in each lane.
To this end, we introduce a remarkably simple and elegant model showcasing the GR universality: a multilane generalization of the symmetric simple exclusion process (SSEP) \cite{Liggett1985} with a Bernoulli stationary measure. As in the SSEP, the model is characterized by only one parameter that merely sets the overall time scale. Thus, like the KPZ and diffusive universality classes, the GR universality class can be realized robustly without fine-tuning or nontrivial microscopic parameters.

\emph{Stochastic systems with circulant symmetry.---}
Consider a Markov process with $M$ conservation laws, whose local densities are denoted by $\rho_\alpha(x,t)$, $\alpha=0,\ldots,M-1$.
In the thermodynamic limit, the dynamics of the conserved densities are described by a system of conservation laws
\begin{align}
\partial_t \rho_\alpha (x, t)
+
\partial_x j_\alpha(\underline{\rho}(x, t))
=0,
\label{eq:continuity}
\end{align}
where $j_\alpha(\underline{\rho}) = j_\alpha(\rho_0,\rho_1,\ldots,\rho_{M-1})$
denotes the stationary current of the conserved quantity $\alpha$ in a homogeneous stationary state characterized by the average densities $\rho_\beta$ of the conserved quantity $\beta$. The flux Jacobian, given by the matrix of the first derivatives with elements 
\begin{align}
&J_{\al \be} = \frac{\partial j_\al}{ \partial \rho_\be},
\label{eq:J}
\end{align}
is  diagonalizable with real eigenvalues $c_\al$, 
\begin{align}
&R J R^{-1} = \mathrm{diag} (c_0,c_1, \ldots ,  c_{M-1}),
\label{eq:Jdiag}
\end{align}
where $R$ is the corresponding transition matrix.
The eigenvalues $c_\al$  represent the characteristic velocities of the normal modes of the system, i.e.~the velocities at which  local perturbations spread on top of a homogeneous stationary background  \cite{2003JStatPhys}.

Suppose now that the Jacobian $J$ is circulant, 
\begin{align}
  &J_{\al+1,\,\be+1} = J_{\al,\be}, \quad \forall\, \al,\be,
  \label{eq:circulant}\\
  &M \equiv 0, \no
\end{align}
where all indices are understood modulo $M$.
Such a symmetry of the Jacobian is guaranteed,  for instance, by the corresponding circulant symmetry of the stationary currents, 
 \begin{align}
&j_{\al+1}(\rho_0, \rho_1, \ldots,  \rho_{M-1})  =  j_{\al}(\rho_1,  \ldots,  \rho_{M-1},\rho_0),  
\label{eq:CurrCirculant}
\end{align}
for all $\alpha$, which in its turn is a direct consequence of the circulant symmetry of the microscopic hopping
rates.  Indeed,  using Eq.~\eqref{eq:CurrCirculant}  and calculating the Jacobian (\ref{eq:J}) for a manifold where  
average densities are all equal,
\begin{align}
&\rho_0 = \rho_1= \ldots = \rho_{M-1} = \rho,
\label{eq:EDscenario}
\end{align}
 one readily finds that Eq.~\eqref{eq:circulant} is automatically satisfied.
As we show below, the global circulant symmetry \eqref{eq:circulant} leads to a robust realization of the Golden ratio
universality class, provided the number of conservation laws $M$ is odd and sufficiently large,
$M \geq 5$.

A circulant matrix is fully specified by its first row
$(J_{0,0}, J_{0,1}, \ldots J_{0,M-1}) = (a_0,a_{1},\ldots,a_{M-1})$,  where all $a_j$ are real by definition of the Jacobian $J$. 
Its eigenvalues are \cite{davis_circulant_1979}
\begin{align}
c_k =a_0 + \sum_{j=1}^{M-1} a_j w^{j k}, \qquad
w = \exp \left(\frac{2 \pi i} { M} \right),
\label{eq:EigCirculant}
\end{align}
with corresponding eigenvectors
\begin{align}
&\vb_{k} = \frac{1}{\sqrt{M}}\left( 1,\, w^{k},\, w^{2k}, \ldots, \, w^{(M-1)k} \right)^T.
\label{eq:EivCirculant}
\end{align}
From Eq.~\eqref{eq:EigCirculant} and $w^M=1$, one obtains $c_\al = c^*_{M-\al}$. Since all the 
characteristic velocities are real, this relation reduces to 
\begin{align}
c_\al = c_{M-\al}.\label{eq:Sym}
\end{align} 
As a consequence, the modes $\alpha$ and $M-\alpha$ propagate with the same velocity,  so that hydrodynamic modes come in degenerate pairs (these are known as \textit{umbilic modes}). The only unpaired (also called \textit{regular} or \textit{singular}) modes are the self-conjugate ones, satisfying $\al = M-\al \mod M$. These correspond to the mode $\alpha = 0$ for any $M$, and additionally $\al = M/2$ when $M$ is even. 

The rows $R_k$, ($k=0,\ldots, M-1$)  of  the diagonalizing matrix $R$ in Eq.~\eqref{eq:Jdiag} are given by $R_k = \vb_k$, while the
rows of the transposed inverse are $(R^{-1})^T_k = \vb_{k}^* = \vb_{-k} $. This follows from the orthogonality relation
\begin{align}
& (\vb_k   \vb_{-j}) = \de_{kj},  \label{eq:orthogonalityFourier}
\end{align}
where indices are understood modulo $M$.

\emph{Mode-coupling analysis.---} We now turn to the mode-coupling matrices $G^\al$,  the central objects in 
the nonlinear fluctuating hydrodynamics (NLFH) framework \cite{Spoohn14}. According to NLFH,
the fluctuations of the normal modes evolve according to the coupled stochastic partial differential equations 
\begin{align}
&\partial_t \phi_\al + \partial_x \left( c_\al \phi_\al
+ G^{\al}_{\be\ga}\, \phi_\be \phi_\ga
- \partial_x D_{\al\be}\, \phi_\be
+ B_{\al\be}\, \xi_\be \right) = 0,
\label{eq:NLFH}
\end{align}
with summation over repeated indices implied.
Here $c_\al$ are the characteristic velocities, $D_{\al \be}$ are the elements of the phenomenological diffusion matrix, and $\xi_\be$ are independent Gaussian white noises with amplitudes fixed by $B$ through the fluctuation--dissipation relation \cite{Spoohn14}. The mode-coupling matrices $G^\al$ with mode-coupling coefficients $G^{\al}_{\be\ga}$ effectively quantify how the quadratic nonlinearity feeds the
pair of modes $(\be,\ga)$ into mode $\al$. By inspecting the vanishing and non-vanishing coefficients one can determine the dynamical universality class of each mode \cite{2015Fibonacci}. 
$G^\al$ are given by 
\beq
G^{\al}= \frac{1}{2} \sum_{\ga} R_{\al \ga} (R^{-1})^T H^{\ga}R^{-1},
\label{eq:Gdef}
\enq
where $H^\al$ represent the Hessians of the currents with elements $H^{\alpha}_{\beta \gamma} = \partial^2 j_{\al}/\partial \rho_{\beta} \partial \rho_{\gamma}$.

The calculation of the mode-coupling matrices $G^\al$ is straightforward.  
Writing the components of the eigenvectors as 
\begin{align}
&v_{j, k} = (\vb_j)_k =\frac{1}{\sqrt{M}} w^{jk},\label{eq:v_jk}
\end{align}
and using the circulant structure of the currents \eqref{eq:CurrCirculant},
which implies that all Hessians follow from $H^0$ by a shift,
$H^\al_{n+\al,\,m+\al} = H^0_{n,m}$, a short calculation yields
\begin{align}
 G^\al_{\be\ga} = \frac{1}{2} \sum_{n,m=0}^{M-1}& H^0_{n,m}
\sum_{k=0}^{M-1} v_{\al,k}\, v_{-\be,\,k+n}\, v_{-\ga,\,k+m},\\
& \al,\be,\ga = 0,\ldots,M-1. \notag
\label{res:G}
\end{align}
The inner sum over $k$ can be carried out explicitly using the Fourier orthogonality relation \eqref{eq:orthogonalityFourier}, giving
\begin{equation}
\begin{aligned}
 G^\al_{\be\ga}
 &= \left(\frac{\sqrt{M}}{2} \sum_{n,m=0}^{M-1}
 H^0_{n,m} v_{-\beta,n}\,v_{-\gamma,m}\right)\\
 &\quad\times \de_{(\be+\ga-\al \bmod M)} .
\end{aligned}
\label{res:GG}
\end{equation}
The structure of Eq.~\eqref{res:GG} is worth emphasizing. The prefactor in brackets is model dependent and contains all information specific to the currents through $H^0$, whereas the selection rule
$\al = \be+\ga \bmod M$ follows solely from the underlying circulant symmetry. In the following we concentrate on the most interesting case of odd $M$, which leads to robust GR universality.

Before proceeding, we recall the conditions under which GR universality
emerges in the case of a nondegenerate Jacobian spectrum (see \cite{SM} for details). A set $\mathcal{A}$ of hydrodynamic modes falls into the GR universality class whenever it is free of self-couplings \eqref{eq:GR1}, internally coupled \eqref{eq:GR2}, and decoupled from the rest of the spectrum \eqref{eq:GR3}. This can be formulated as follows,
{\tagsleft
\begin{align}
& \qquad G^{\al}_{\al\al} = 0,
\tag{GR1}\label{eq:GR1}\\[2pt]
& \qquad G^{\al}_{\be\be} \neq 0 \;\text{for some } \be\neq \al
\text{ in } \mathcal{A},
\tag{GR2}\label{eq:GR2}\\[2pt]
& \qquad G^{\al}_{\ga\ga} = 0 \;\text{for } \ga\notin\mathcal{A},
\tag{GR3}\label{eq:GR3}
\end{align}
for all $\alpha \in \mathcal{A}$.
}

For a doubly degenerate (umbilic) mode $(\al,M-\al)$, cf.~Eq.~\eqref{eq:Sym}, the notion of self-coupling in the NLFH equations \eqref{eq:NLFH} must be generalized. In particular, the self-coupling of the umbilic mode $\al$ is determined by the  eight coefficients $G^{j}_{k,l}$ with $j,k,l=\{ \al,M-\al\}$.
The self-coupling vanishes  when \textit{all} of them vanish.  Similarly,  the coupling between two distinct umbilic modes $\al$ and $\ga$ is given by eight matrix elements $G^{\al}_{k,l}$,$ G^{M-\al}_{k,l}$ for $k,l=\{ \ga,M-\ga \}$. Finally, the coupling of an umbilic mode $\al$ to the regular mode $0$ (or $M/2$, if $M$ is even) is characterized by the two coefficients, $G^\al_{0,0},G^{M-\al}_{0,0}$.  In the following analysis, we will assume that the model-dependent prefactor in Eq.~\eqref{res:GG} is nonzero.
The mode-coupling structure thus depends solely on the Kronecker $\delta$ selection rule of Eq.~\eqref{res:GG}. We will furthermore assume that no additional degeneracies occur beyond those implied by Eq.~\eqref{eq:Sym}. Under these assumptions, we summarize below the universality classes predicted by MCT along the equal-density line (\ref{eq:EDscenario}).

\textbf{Case $M=3$.}
There are two modes, one regular, $\alpha=0$ with velocity $c_0$, and one doubly
degenerate umbilic mode with velocity $c_1=c_2$.
Both modes generically have nontrivial self-coupling,   
$G^{0}_{0,0} \times G^{1}_{2,2}\times G^{2}_{1,1} \neq 0$. Due to this, MCT predicts the same dynamical exponents, $z_0=z_1=3/2$. However, the modes have different scaling functions: the regular mode has the usual Pr\"ahofer-Spohn scaling function of the KPZ universality class \cite{Prae04}, while the umbilic mode is characterized by the recently discovered KPZ-like  scaling function for a doubly degenerate umbilic mode with self coupling \cite{2024Spohn,2026Spohn, 2026UmbilicMultilane}.  For odd $M>3$ the situation is qualitatively different,  as discussed below.

\textbf{Case $M=5,7,9,\ldots, \infty$.}
For odd $M \geq 5$ the model has one regular nondegenerate mode, while the 
remaining modes  are
doubly degenerate, 
see Eq.~\eqref{eq:Sym}.
From Eq.~\eqref{res:GG} we find the following (see \cite{SM} for details),
\begin{enumerate}[label=(\roman*)]
    \item Every umbilic mode $\al\neq M/3$ is coupled to another umbilic mode $\be \neq M/3$,   and is not coupled to itself.  A mode with $\al=M/3$ (only present for odd $M=9,15,\ldots$ divisible by $3$) is self-coupled.
    \item Umbilic modes are not coupled to the regular mode. 
    \item The regular mode is coupled to itself and to all umbilic modes.
\end{enumerate}

It follows 
that  all the umbilic modes, except for $\al=M/3$, belong to the GR universality class.
We emphasize once again that this result follows solely from the circulant symmetry \eqref{eq:CurrCirculant}.
The remaining modes have the KPZ dynamical exponent,  $z_0=
z_{M/3}=3/2$. 

If,  in addition, the self-coupling of the regular mode vanishes, c.f.~Eq.~(\ref{res:GG}),
 \begin{align}
&  G^0_{00} = 
 \frac{1}{2\sqrt{M}} \sum_{n,m=0}^{M-1} H^0_{n,m} =0,
\label{res:G000}
\end{align}
then also the regular mode $\al=0$ showcases GR universality. The constraint \eqref{res:G000} has a transparent physical meaning, as it corresponds to a vanishing curvature 
of the function $j_0(\rho,\rho, \ldots, \rho)$ in the direction $(1,1,\ldots , 1)$. 
Indeed, suppose 
\begin{align}
& j_0(\rho,\rho, \ldots, \rho)\equiv 0, \quad \forall \rho.
\label{res:G000meaning}
\end{align}
Expanding the difference 
$j_0(\rho+\de \rho , \ldots, \rho+\de \rho)  -j_0(\rho,\rho, \ldots, \rho)$ in a Taylor series,  we obtain Eq.~\eqref{res:G000} in the second-order
of the expansion.

\textbf{Case $M$ even.}
For $M$ even,  generalized MCT
does not predict GR universality for any subset of modes.  Instead,  robust Fibonacci universality classes 
\cite{2015Fibonacci} with 
$z_\al = F_{n}/F_{n-1}$ are predicted; however, they require exponentially growing $M= O( 2^n)$: details to be 
found in  \cite{Longpaper}.

\emph{Multilane symmetric exclusion process.---} To verify our theoretical predictions we introduce a family of stochastic models with a circulant symmetry, which, on the GR manifold \eqref{eq:EDscenario}, can be viewed as  a multilane generalization of the symmetric simple exclusion process (SSEP).

Consider $M$ identical parallel lanes of $L$ sites with periodic boundary conditions in both directions, $L+1 \equiv 1$ and $M+1=1$. 
Each lane is a one-dimensional lattice with sites that can be either occupied by a particle or be empty.  A particle can hop bidirectionally to the nearest neighbor site on the same lane when the target site is empty.  Hoppings between lanes are forbidden,  yielding particle conservation in each lane,  leading to a system of $M$ conservation laws \eqref{eq:continuity} in the hydrodynamic limit.
 
\begin{figure}[tbp]
\centerline{
\includegraphics[width=\columnwidth]{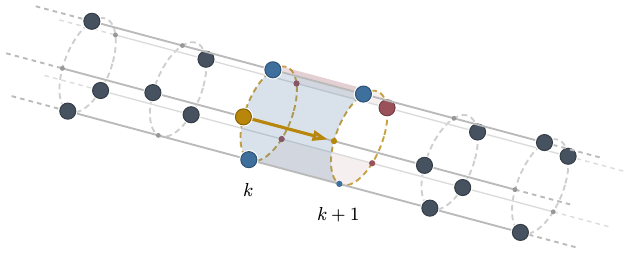}
}
\caption{ Schematic setup of the $M$-lane SSEP (\ref{eq:RatesBernouilly}) with $M=5$.  The
circulant symmetry of the rates is realized by the invariance of the rates under the rotation of the tube by the angle $2\pi/M$.
Particles (filled disks) hop along a lane and never hop across,  ensuring particle number conservation in each lane. 
The net drift  of the chosen yellow particle (\ref{eq:RatesBernouilly}) depends on the
occupations of the tinted surface only:   particles  at lateral
distance $s=1$ (coloured blue) and $s=2$ (coloured wine) enter with weights $\ga_1$,  and  weights $-\ga_1$
respectively.
  The shown configuration contains three blue and one wine particle,  resulting
in the net right drift of the chosen particle 
\mbox{$p^\lambda_{k,k+1}-q^\lambda_{k,k+1} = (\ga_1/2)(3-1) = \ga_1$}.   
}
\label{Fig5lanes}
\end{figure}

The hopping rates depend on the occupation of the neighboring sites in the adjacent lanes, which introduces  interactions between the lanes.  The dynamics are governed by a continuous time master equation.  Denoting the right and left hopping rate of a particle  on lane $\mu$ at the link $(k,k+1)$  as
 $p_{k,k+1}^\mu$ and $q_{k,k+1}^\mu$ respectively,  we postulate their difference to be
 \begin{align}
& p_{k,k+1}^\mu - q_{k,k+1}^{\mu} =\sum_{s=1}^{\frac{(M-1)}{2}}  
\frac{\gamma_{s}}{2} \sum_{\ka=\pm s} \left( n_{k}^{\mu+\ka} + n_{k+1}^{\mu+\ka} \right),
\label{eq:RatesBernouilly}
\end{align}
where $n_{k}^{\mu}=0,1$ is a particle occupation number on site $k$ on lane $\mu$,
and  $\ga_s$ represent the interlayer coupling amplitudes  between a lane $\mu$ and  the lanes at a lateral distance $s$ from it  $\mu\pm s$.  Note that the rates depend on sums of occupations  
  of type $\left( n_{k}^{\nu} + n_{k+1}^{\nu} \right)$ and not on individual  occupations $n_{k}^{\nu}, n_{k+1}^{\nu} $. 
If other lanes are empty ($n_{k}^{\nu}=0$,  $\forall k,  \ \forall \nu \neq \mu$ ),  the dynamics  on lane $\mu$
reduces to 
the symmetric simple exclusion process (SSEP).  
For five lanes,  the Markov process with rates (\ref{eq:RatesBernouilly}) is illustrated in Fig.~\ref{Fig5lanes}.  

The continuous time Markov process with rates \eqref{eq:RatesBernouilly}
on an infinite lattice has a Bernoulli product stationary measure (see \cite{SM}) which allows to easily obtain the 
stationary current-density relation,
\begin{align}
&j_{\mu}(\underline{\rho}) = \rho_{\mu} (1 - \rho_\mu)  
\sum_{s=1}^{(M-1)/2}  
\gamma_{s}  \sum_{\ka=\pm s}   \rho_{\mu +\ka},
\label{eq:Curr}
\end{align}
where $\rho_{\mu} = \langle n_{k}^{\mu} \rangle$ is the average particle density on lane $\mu$.
With $M+s\equiv s$, the current-density relation \eqref{eq:Curr} satisfies the
circulant condition \eqref{eq:CurrCirculant},  guaranteeing the existence of the GR subset of umbilic modes
for odd $M>3$ 
and equal particle densities (\ref{eq:EDscenario}). 
Moreover,  if
\begin{align}
&\sum_{s=1}^{(M-1)/2}\ga_s=0, \label{eq:SumGammaZero}
\end{align} 
then  net steady drift of particles on all lanes vanishes, 
$j_\mu(\rho,\ldots,\rho)\equiv 0$, justifying the  name multilane symmetric exclusion process.  The conditions \eqref{res:G000} and \eqref{res:G000meaning}
follow, causing also the regular mode to belong to the GR universality class.  
For any odd $M>3$,  and for any choice of constants $\ga_s$ satisfying Eq.~\eqref{eq:SumGammaZero} we predict all the conserved modes (except mode $\al=M/3$) to showcase GR universality.

\emph{Minimal GR model. ---}
 For the simulations we take $M=5$, the smallest lane number for which  the relations 
\eqref{res:GG} and  (\ref{eq:SumGammaZero})  ensure the emergence of GR universality for
all modes, both regular and umbilic, for equal particle densities on all lanes $\rho_\al \equiv \rho$.

The resulting minimal GR model is illustrated in Fig.~\ref{Fig5lanes}.  It has no
 parameters,  apart from $\ga_1= -\ga_2 \neq 0$ determining the overall 
time scale,  like in a usual  single-lane SSEP.  

\begin{figure}[tbp]
    \centering
\includegraphics[width=\columnwidth]{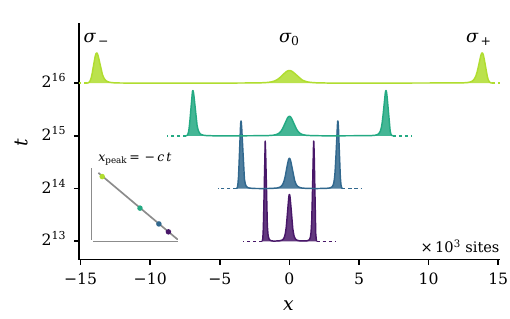}
    \caption{
        {\bf Space--time propagation of the three normal modes Eq.~(\ref{Scalingfunction}) in real space.}
        Monte Carlo structure factor of the $M=5$ multilane SSEP, at different times. 
        The regular mode $\si_0$ stays at the origin while the two umbilic modes $\si_\mp$ travel at
        $c_\mp = \mp \frac{\sqrt{5}}{2}\rho(1-\rho)$. Each mode is a
        L\'evy packet centered on $c_\al t$, of width $\propto t^{1/\vfi}$ and asymmetry $\be=1,0,-1$,
see also Fig.~\ref{fig:M5_NNN} for a closeup.  Note that the time  lags are powers of two stacked at equal
        spacing, so the fan opens exponentially across the stack, whereas the drift itself is
        ballistic.   The inset shows the umbilic peak position on linear axes, a straight line of
        slope $c$ through the origin.  Simulation parameters are as in Fig.~\ref{fig:M5_NNN}, and both Fig.s~\ref{fig:propagation},\ref{fig:M5_NNN} have the same colour code.
}
    \label{fig:propagation}
\end{figure}

The minimal model has two doubly degenerate umbilic modes coupled to each other  and one regular mode coupled to both umbilic modes.
There are no self-couplings.
The characteristic velocities  are arranged symmetrically: the regular mode is immobile,  $c_0=0$, while the velocities of the two 
umbilic modes are $\pm  \frac{\sqrt{5}}{2} \rho(1-\rho)$.  A point-like excitation thus spreads symmetrically,  see Fig.~\ref{fig:propagation}. All the nonzero couplings  are proportional to a single parameter, $(1-2\rho)$,  which should remain nonzero, i.e.~$\rho \neq 1/2$ (see \cite{SM}).  Generalized MCT predicts GR for all three modes \cite{SM}, with scaling functions of the same type, but with different asymmetry $\be$.
The umbilic modes have maximally asymmetric $\be= \pm 1$ Lévy stable GR scaling functions related by a mirror symmetry, and the regular mode has a symmetric $\be= 0$ Lévy stable GR profile.  
The Lévy stable GR scaling functions are probability distribution functions which in momentum space have the form \cite{Durrett}
\begin{align} \label{Scalingfunction}
&\hat{S}_\alpha(p,t) = \frac{1}{\sqrt{2\pi}} \rme^{-ic_\alpha
  pt-E_\alpha |p|^{z}t \left(1 - i \, 
   \mathrm{sgn}(p) \be\tan{\left(\frac{\pi z}{2}\right)}
  \right)}
\end{align}
with $z\equiv \varphi$.

\begin{figure*}[tbp]
    \centering
\includegraphics[width=\textwidth]{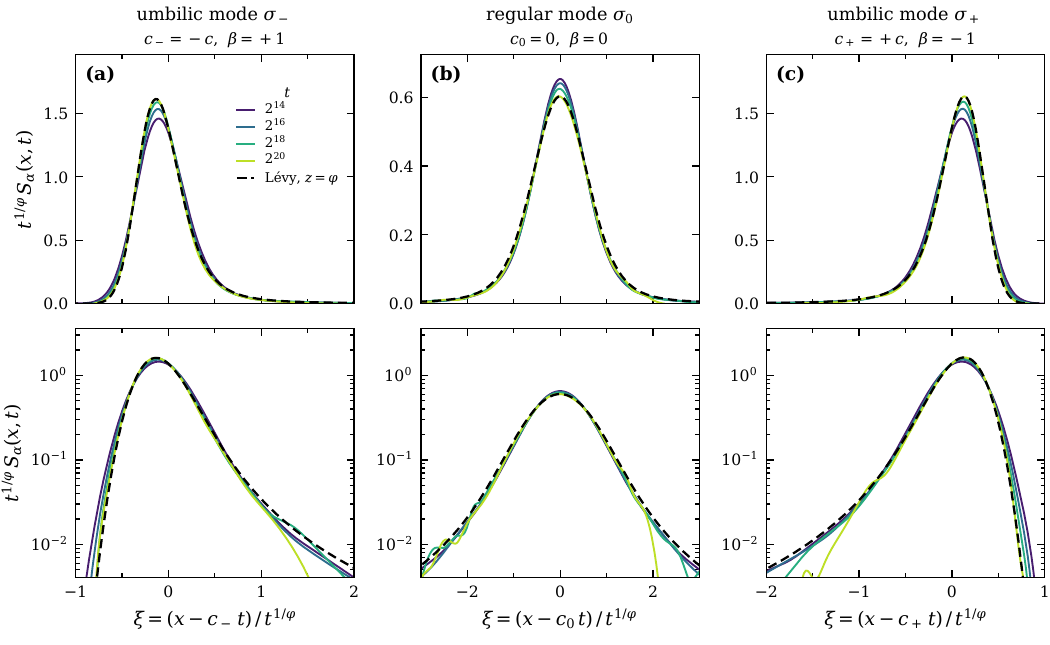}
    \caption{
        {\bf Dynamic structure factor for all modes of the $M=5$ multilane SSEP with Bernoulli rates.}
        Data collapse with the predicted GR
        dynamical exponent $z=\vfi$ of the comoving structure factor $S_\al(x,t)$ at times $t = 2^{14}, 2^{16}, 2^{18}, 2^{20}$.
Monte Carlo simulation data are given by full lines,
see Fig.~\ref{fig:propagation} for colour coding.  The panels are ordered by characteristic
        velocity. The left (a) and right (c) structure factors belong to umbilic modes and are related by  a mirror symmetry.  
The lower row shows the data in logarithmic scale for the $y$ axis. The black dashed curve is the L\'evy-stable
        scaling function (\ref{Scalingfunction}) with $z=\vfi$ and $\be=+1$ (a), $\be=0$ (b),
        $\be=-1$ (c); its position and scale are pinned to the measured peak, so it carries no free shape parameter.  Simulation parameters: $M=5$,
        $L=2^{20}$, $\rho=1/4$, $\ga_1 = -\ga_2=1/2$, $N_{\mathrm{samples}}=10^3$. 
}
    \label{fig:M5_NNN}
\end{figure*}

The results of the Monte Carlo simulations for the multilane $5$-lane SSEP with rates \eqref{eq:RatesBernouilly}  are in full agreement with the above generalized MCT predictions,  see  Figs.~\ref{fig:propagation},~\ref{fig:M5_NNN}. 
The dynamic structure factor of the regular mode converges rapidly to the predicted asymptotic form \eqref{Scalingfunction}, whereas convergence for the umbilic modes is considerably slower.

\emph{Discussion. ---} To conclude,  we proposed 
a robust way to realize the Golden ratio universality class in 
stochastic many-body systems with conservation laws, arising from symmetry rather than by fine tuning. 

The key condition, a \textit{circulant} symmetry of the currents in the steady state, is conveniently realized in multilane Markov processes with particle number conservation in each lane and rates that are translationally invariant in the transverse (lane) direction.  Remarkably, the GR universality class arises here in weakly hyperbolic systems with doubly degenerate (umbilic) conserved modes.  Parity is crucial: 
given the circulant symmetry, an \textit{odd} number of conservation laws, $M\geq 5$, is required to observe GR universality. Finally, we proposed a multilane version of the symmetric simple exclusion process with Bernoulli stationary measure. This is a minimal,  parameter-free model that exhibits GR universality, as confirmed by Monte Carlo simulations. 

Our findings establish the GR universality class as a robust feature of many-body systems with a circulant symmetry, rather than an exotic prediction of mode-coupling theory.

\begin{acknowledgments}
L.~P.~thanks Friedrich H\"ubner, Enej Ilievski, Matija Koterle, and \v{Z}iga Krajnik for insightful discussions. The authors thank Enej Ilievski and Matija Koterle for useful comments on the manuscript.
L.~P.~acknowledges support from the European Union HORIZON-CL4-2022-QUANTUM-02-SGA through PASQuanS2.1 (Grant Agreement No. 101113690), as well as the Slovenian Research and Innovation agency (ARIS) through the Program P1-0402 and Grant N1-0368.
V.P.  acknowledges support by ERC Advanced grant
 No.~101096208 -- QUEST,  Research Program P1-0402 and Grant N1-0368 of Slovenian Research and Innovation Agency (ARIS) and by Deutsche Forschungsgemeinschaft through DFG project KL645/20-2. 
\end{acknowledgments}

\bibliographystyle{apsrev4-2}
\bibliography{references}

\clearpage

\setcounter{table}{0}
\renewcommand{\thetable}{S\arabic{table}}%
\setcounter{figure}{0}
\renewcommand{\thefigure}{SM\arabic{figure}}%
\setcounter{equation}{0}
\renewcommand{\theequation}{S\arabic{equation}}%
\setcounter{page}{1}
\renewcommand{\thepage}{SM-\arabic{page}}%
\setcounter{secnumdepth}{3}
\setcounter{section}{0}
\renewcommand{\thesection}{\arabic{section}}%
\setcounter{subsection}{0}
\renewcommand{\thesubsection}{\arabic{section}.\arabic{subsection}}%
\renewcommand{\thesection}{S-\Roman{section}}

\onecolumngrid

\global\long\def\no{\nonumber}

\begin{center}\Large{
		\textit{Supplemental Material for}\\
		\medskip
		
		\textbf{Robust Golden Ratio Universality as a consequence of a circulant symmetry}\\
		
		\medskip
		by Luka Paljk,  Johannes Schmidt and Vladislav Popkov
	}
\end{center}

{
\hypersetup{linkcolor=black}
}

The Supplemental Material contains four sections. 
In App.~\ref{app:Bernoulli} we prove that the minimal GR model admits a Bernoulli stationary measure. 
In App.~\ref{app:GRconditions} we recall the conditions on the mode-coupling matrices that give rise to Golden Ratio universality.
In App.~\ref{app:GRsubsetproof} we prove the Golden Ratio scenario for an odd number of conservation laws in systems with circulant symmetry.
Finally, in Appendix \ref{app:MCT}, we briefly outline how mode-coupling theory can be generalized to treat degenerate modes.

\section{Bernoulli stationary measure for the Markov process with rates (\ref{eq:RatesBernouilly})}
\label{app:Bernoulli}
To prove that the Markov process with rates \eqref{eq:RatesBernouilly} has a Bernoulli stationary measure, we take a system with $M$ lanes, each lane having $N$ sites, with periodic boundary conditions $N+k \equiv k$. 

Assuming  fixed total particle number in each lane,  and uniformity of a unique stationary measure,  i.e.~that the stationary state probabilities $P(C)$ of all configurations are equally likely, 
\begin{align}
&P(C)=P(C'), \quad \forall C,C'. \label{eq:Bernoulli}
\end{align}
we check (\ref{eq:Bernoulli}) by verifying the  stationary master equation
\begin{align}
&\sum_{C'\neq C} w_{C',C} P(C') = P(C) \sum_{C''\neq C} w_{C,C''}   \label{SteadyW}
\end{align}
which,  using (\ref{eq:Bernoulli}), becomes
\begin{align}
&  \sum_{C''\neq C} w_{C,C''}- \sum_{C'\neq C} w_{C',C} =0. \label{SteadyBernoulli}
\end{align}
Here $C$ is an arbitrary configuration, and $w_{C,C'}$ is a rate of hopping from configuration $C \rightarrow C'$.

For the proof we take a slightly more general rate difference than Eq.~\eqref{eq:RatesBernouilly}. We consider a difference $r$ between the right and left hopping rates on lane $\al$ on the link $k,k+1$ of the form
\begin{align}
 r^\al_{k,k+1}\equiv  p^\al_{k,k+1} - q^\al_{k,k+1}&= B_\al + \sum_{\mu \neq \al}  
\frac{g(\al, \mu)}{2}  \left( n_{k}^{\mu} + n_{k+1}^{\mu} \right),
\label{eq:RatesBernouillyGeneral}\\
g(\al ,\mu)&=g(\mu,\al).\label{symmetricTensor}
\end{align}
Denoting the occupation of site $k$ on lane $\al$ as $n^\al_{k}$,
the left-hand side of Eq.~\eqref{SteadyBernoulli} is
equal to 
\begin{align}
&  \sum_{C''\neq C} r_{C,C''}- \sum_{C'\neq C} r_{C',C} =\sum_\al \sum_k \left( 
n^\al_{k}- n^\al_{k+1} 
\right)  r^\al_{k,k+1}.
\end{align}
Substituting this into Eq.~\eqref{eq:RatesBernouillyGeneral}, and by  noting that $\sum_\al \sum_{\mu \neq \al} Z_{\al \mu} = 
\sum_{\mu>\al} (Z_{\al \mu}+Z_{\mu \al} )$, we obtain
\begin{align}
&\sum_\al \sum_k \left( 
n^\al_{k}- n^\al_{k+1} 
\right)  r^\al_{k,k+1} = 
\sum_{k} \left( 
\Phi_{k}- \Phi_{k+1}
\right),  \label{telescope}  \\
& \Phi_{k}= \sum_\al B_\al n^\al_{k} + \sum_{\mu > \al} g(\al, \mu) n^\al_{k} n^\mu_{k}\no,
\end{align}
where we also used the relation \eqref{symmetricTensor}. On a periodic lattice, the sum \ref{telescope} telescopes to zero,  proving stationarity of a uniform canonical measure (\ref{app:Bernoulli}), and stationarity of the grand-canonical Bernoulli product measure for the rates (\ref{eq:RatesBernouillyGeneral}).
The special case \eqref{eq:RatesBernouilly} follows from \eqref{eq:RatesBernouillyGeneral} by choosing 
$B_\mu \equiv 0$,  and
$\ga_s = g(\mu, \mu \pm s \mod M)$,  $\forall \mu$.

\section{Golden ratio universality conditions in systems with several conservation laws}
\label{app:GRconditions}

It was shown in Ref.~\cite{2015Fibonacci} that a self-consistent analysis of 
the mode coupling equations for hyperbolic systems with several conservation laws
leads to a prediction of different universality classes, in particular dynamical exponents $z_\al$ characterizing the
relaxation of the perturbations of conserved modes on top of a stationary homogeneous background.
The MCT predictions for mode $\al$ are based on the  set $\I_\alpha := \{\beta:
G^{\alpha}_{\beta\beta} \neq 0\}$ of non-zero diagonal coefficients of the mode-coupling matrix $G^\al$.
One obtains a system of equations for the dynamical exponents,
\begin{equation}
\label{app:DynExponents}
z_\alpha = \left\{ \ba{lll}
2 & \mbox{ if } & \I_\alpha = \emptyset  \\
3/2 & \mbox{ if } & \alpha \in \I_\alpha\\
\min_{\beta \in \I_\alpha}\left[\left(1 + \frac{1}{z_\beta}\right)\right]
& \mbox{ else } &
\ea \right.
\end{equation}
which should be solved  self-consistently for all $\al$.

The GR scenario for all modes arises if 
\begin{align}
&G^\al_{\al \al} =0,\quad \forall \al,    \label{app:NoSelf} \\
& \rm{diag} (G^\al)  \neq 0, \quad  \forall \al  \label{app:GR_diag}
\end{align}
For a proof, we write the system of equations according to (\ref{app:DynExponents}): 
\begin{align}
&z_\al =\min_{\beta \in \I_\alpha}\left[\left(1 + \frac{1}{z_\beta}\right)\right], \quad \forall \al.
\end{align}
The only self-consistent solution of the above equation is 
\begin{align}
&z_\al =z,\quad z=1+ \frac{1}{z},
\end{align}
i.e.~GR universality for all modes
\begin{align}
&z_\al\equiv \varphi =\frac{1+ \sqrt{5}}{2}, \quad \forall \al.
\end{align}
In words, the GR conditions \eqref{app:NoSelf} and \eqref{app:GR_diag} amount to 
\begin{align}
&   \mbox{-- \textit{no self couplings}}\no\\
&\mbox{--  each  mode \textit{is coupled} to some other mode/modes. } \label{app:GR}
\end{align}
A subset $\mathcal{B} \subset U$ where $U$ is the full set of modes thus belongs to the GR universality class if 
\begin{align}
& \alpha \notin \I_\alpha, \quad  I_\al \cap \mathcal{B} \neq \varnothing,  \quad I_\al \cap (U \setminus \mathcal{B})   =  \varnothing  \quad  \forall \al \subset \mathcal{B},
\end{align}
A straightforward analysis of the above conditions according to the prescription given by Eq.~\eqref{app:DynExponents}
leads to \eqref{app:GRsubset}.

\section{Proof of umbilic GR subset for odd $M\geq  5$}
\label{app:GRsubsetproof}
We consider systems with circulant symmetry \eqref{eq:CurrCirculant} presented in the Main Text.
For odd $M$, we have one regular mode $\al=0$ and $(M-1)/2$ doubly degenerate modes,  with $\al=1,2, \ldots , (M-1)/2$.  Below we prove the scenario in which GR universality emerges (\ref{app:GRsubset}). For a subset $\mathcal{B}$ of doubly degenerate (umbilic) modes  $\{\al\} =\{1,2, \ldots , (M-1)/2 \}$ we require the following.
\begin{align}
&   \mbox{ (i) \textit{No self couplings} within $\mathcal{B}$.}\no\\
&\mbox{ (ii)  Each mode in $\mathcal{B}$ \textit{is coupled} to some other $\mathcal{B}$ modes.}  \label{app:GRsubset} \\
&\mbox{  (iii) Modes inside of $\mathcal{B}$  are  \textit{not coupled} to modes outside of the subset.} \no
\end{align}

(i) All the self-coupling coefficients of an umbilic mode $\al$ vanish  $\forall \al$.\\
To establish this property, we need to show that all 8 matrix elements $G^{j}_{k,l}$ with $j,k,l=\{ \al,M-\al\}$ are zero. We rewrite Eq.~\eqref{res:GG} as $G^{j}_{k,l}= f(j,k,l) \,K^{j}_{k,l}$,
where $f(j,k,l)$ is a model-dependent term, and the
Kronecker delta factor 
\begin{align*}
&K^{j}_{k,l} =K^{j}_{l,k} =\de_{(k+l-j \mod M)}
\end{align*}
is universal. 
Furthermore, we will assume the model-dependent term to be generically nonzero for all $j,k,l$, focusing solely on the selection rule given by $K^{j}_{k,l}$.
Analysing the universal Kronecker delta factor for $0<\al\leq (M-1)/2$, gives us
\begin{align}
&K^\al_{\al, \al} = K^{M-\al}_{M-\al, \al} =  K^{M-\al}_{\al, M-\al} = \de_{(\al \mod M)}=0, \no\\
&K^{M-\al}_{M-\al, M-\al} = K^{\al}_{M-\al, \al} =  K^{\al}_{\al, M-\al} = \de_{(M-\al \mod M)}=0, \no\\
&K^\al_{M-\al, M-\al} =  \de_{(-3\al \mod M)}=0,  \quad \mbox{if } \ \frac{M}{3} \notin \mathbb{N}, \label{Mdiv3exception} \\
&K^{M-\al}_{\al, \al} =  \de_{(3\al \mod M)}=0, \quad \quad \quad  \quad \mbox{if } \ \frac{M}{3} \notin \mathbb{N}.  \no
\end{align}

(ii) Two different umbilic modes $\al, \ga$ are coupled.\\
From the selection rule, we get
\begin{align}
&K^\al_{\ga, \ga} =  K^{M-\al}_{M-\ga, M-\ga} = \de_{(2\ga -\al \mod M)}=1 \quad \mbox{iff $\ga = \al/2 $}, \no\\
&K^\al_{M-\ga, M-\ga} =  K^{M-\al}_{\ga, \ga} = \de_{(2\ga +\al \mod M)} =1, \quad \mbox{iff $\ga = (M-\al)/2 $},
\label{UmbilicCoupling}\\
&K^\al_{M-\ga, \ga} =  K^{M-\al}_{M-\ga, \ga} =K^\al_{\ga,M-\ga} =  K^{M-\al}_{\ga,M-\ga}= 0.\no
\end{align}
Consequently,  every umbilic mode $\al$ is  coupled to another umbilic mode $\ga\neq \al$ : $\ga=\al/2$ for $\al$ even and $\ga= (M-\al)/2 $ for 
$\al$ odd.

(iii) No coupling of umbilic modes to a mode outside of $\mathcal{B}$.\\
Umbilic modes are not coupled to the regular mode $\alpha = 0$, since
\begin{align}
&K^\al_{0, 0} =  K^{M-\al}_{0,0} = 0  \quad \forall \al>0.  \label{NoOutsideCoupling}
\end{align}

From \eqref{Mdiv3exception}-\eqref{NoOutsideCoupling} it follows that for odd $M\geq 5$ non-divisible by $3$, the
GR subset conditions \eqref{app:GRsubset} are satisfied for the whole set of umbilic modes 
$\mathbf{U} := \{1,2,\ldots ,  (M-1 )/2\}$.

Furthermore, for $f(j, k, l)\neq 0$, the single mode is self-coupled and coupled to all other modes:
\begin{align}
&K^0_{0, 0} = 1, \no \\ 
 &K^0_{\be,M-\be} = K^0_{M-\be,\be} =1,  \quad \forall \be=1,2,\ldots ,(M-1)/2.
\label{zeroModeCoupling}
\end{align}

\subsection{Case odd $M\geq 5$ divisible by $3$}
A straightforward check shows that for odd $M\geq 5$ divisible by $3$,  the GR subset still exists,  consisting 
of all umbilic modes but one, $ \mathbf{U} \setminus (M/3)$.  Indeed,  we find that while the mode $\be = M/3$ is self-coupled, 
see Eq.~\eqref{Mdiv3exception},  no other umbilic mode $\al \neq M/3$ is coupled to the mode $M/3$.  
As a result,  the mode  $\be = M/3$ is a doubly degenerate KPZ-like mode with a scaling function found recently in Ref.s~\cite{2025Spohn,2026UmbilicMultilane}, different than the KPZ scaling function of Pr\"ahofer and Spohn \cite{Prae04}. All other umbilic modes 
$\al >0,  \al \neq M/3$ belong to the GR subset.  The regular mode $\al=0$ is a KPZ mode with  Pr\"ahofer-Spohn scaling function \cite{Prae04}. 

\section{Generalized mode coupling theory}
\label{app:MCT}

In a system with $M$ conservation laws, the usual mode-coupling equations for the diagonal entries of the dynamical structure factor 
$S_{\alpha \beta}(x,t) =\exval{\phi_\alpha(x,t)\phi_\beta(0,0)}$ are \cite{Spoohn14}
\begin{equation}
\begin{split}
\partial_t S_{\al\be}(x,t) ={}& \big(-v_\al \partial_x + \textstyle\sum_\ga D_{\al\ga} \partial_x^2\big)\, S_{\ga\be}(x,t) \\
&+ \int_0^t \rmd s \int_{\R} \rmd y \; \partial_y^2 \sum_\ga M_{\al\ga}(y,s)\, S_{\ga\be}(x-y,\, t-s),
\end{split}
\qquad \al,\be = 0, 1, \ldots, M-1,
\label{modecoupling0}
\end{equation}
\begin{align}
M_{\alpha\gamma}(y,s) = 2 \sum_{\mu,\nu,\mu',\nu'=0}^{M-1}
\dG^{\alpha}_{\mu\nu}\,\dG^{\gamma}_{\mu'\nu'}\,
S_{\mu\mu'}(y,s)\,S_{\nu\nu'}(y,s).
\label{app:MemoryKernel}
\end{align}
Here $\dG^{\alpha}$ are canonical mode coupling matrices calculated from \eqref{eq:Gdef} using the
canonical form of $R$ matrices (cf.~Eq.~\eqref{eq:Gdef}) satisfying the relations
\begin{align}
\mathcal{R} J \mathcal{R}^{-1} = \mathrm{diag}(c_0,\ldots,c_{M-1}),  \quad \mathcal{R} C \mathcal{R}^{T} = \mathds{1},
\label{app:canonicalR}
\end{align}
where $C$ is the compressibility
matrix \cite{Spoohn14}. 
The canonical $\mathcal{R}$  and unitary $R$ with elements $R_{\al \be}=M^{-1/2}\exp \left(\frac{2 \pi i \al \be} { M} \right) $ 
are related by $R = W \mathcal{R}$, where
$W$ is unitary and block diagonal,
\begin{align}
W_{00}=1,\qquad
\begin{pmatrix} W_{\mu\mu} & W_{\mu,M-\mu}\\ W_{M-\mu,\mu} & W_{M-\mu,M-\mu}\end{pmatrix}
=\frac{1}{\sqrt2}\begin{pmatrix}1 & i\\ 1 & -i\end{pmatrix}, \qquad \mu = 1,\ldots,(M-1)/2.
\end{align}
Consequently,  the elements of $\dG^{\alpha}$ needed for the mode coupling analysis are related to the elements of
(\ref{res:GG}) via
\begin{align}
\dG^{\alpha}_{\beta\gamma} = \sum_{\lambda,\mu,\nu} W^*_{\lambda\alpha}\, W_{\mu\beta}\, W_{\nu\gamma}\;
G^{\lambda}_{\mu\nu},
\label{eq:GcanG}
\end{align}
with $\alpha\neq 0$, $\gamma=\beta$ for the diagonal elements and $\gamma = M-\beta$ for the off-diagonal element
within a degenerate pair.

To proceed,  we need to establish some crucial properties of  $S_{\alpha \be}$ from (\ref{modecoupling0}).
The ``bare'' fluctuation fields
\begin{align}
s_{\lambda\mu}(k,t)
= \left\langle
\bigl(n^\lambda_{k,t}-\rho\bigr)
\bigl(n^\mu_{0,0}-\rho\bigr)
\right\rangle \no
\end{align}
satisfy the  circulant symmetry
 \begin{align}
&s_{\lambda\mu} = s_{\lambda+1,\mu+1},\label{app:Circulant}
\end{align}
and,  due to invariance of  the rates \eqref{eq:RatesBernouilly}  under lane reflection, $\lambda \to -\lambda$, 
the mirror symmetry relation
\begin{align}
s_{\lambda\mu}(k,t)=s_{\mu\lambda}(k,t),
\qquad \forall \lambda,\mu.
\label{app:MirrorSym}
\end{align}
Due to the properties (\ref{app:Circulant}) and \eqref{app:MirrorSym},
\beq
s_{\lambda\mu} = f(\lambda-\mu), \quad  f(x)=f(-x).
\label{eq:lanedifference}
\enq
Equation \eqref{eq:lanedifference} implies two crucial properties of the dynamic structure factor matrix elements corresponding to the umbilic modes 
$\al = 1, \ldots, (M-1)/2$:
\begin{align}
(i) \qquad & S_{\al,\al}(x,t) = S_{M-\al,M-\al}(x,t), \label{app:Ssym}\\
(ii)\qquad & S_{\al,M-\al}(x,t) = -S_{M-\al,\al}(x,t) = 0. \label{app:antisymmetry}
\end{align}
Both follow from the relation between $S_{\al \be}$ and $s_{\la \mu}$
\beq
S_{\be,\ga}
= \sum_{\lambda,\nu} \mathcal{R}_{\be\lambda}\mathcal{R}_{\ga\nu}\, s_{\lambda\nu}
\enq
and from the explicit form of the canonical matrices $\mathcal{R}$ with the umbilic components
\beq
\mathcal{R}_{\al \nu} = \sqrt{\frac{2}{M}}\cos\left(\frac{2\pi\al \nu}{M} \right), \qquad
\mathcal{R}_{M-\al,\nu} = \sqrt{\frac{2}{M}}\sin\left(\frac{2\pi\al \nu}{M} \right),
\label{eq:Rows_canonical}
\enq
after some tedious but straightforward calculation.

\subsection{Effective mode coupling equations}
Since overlaps $S_{\ga\be}$ for modes with different characteristic velocities $c_\ga \neq c_\be$  eventually die out, 
for the asymptotic analysis  the sum over $\ga$ in (\ref{modecoupling0}) for $\partial_t S_{\alpha \al}(x,t)$
contains a single element 
$ M_{\alpha\al}(y,s)  S_{\al\al}(x-y,t-s)$ \cite{Spoohn14}.  
Denoting $S_\mu(x,t) \equiv S_{\mu \mu}(x,t)$ and accounting for (\ref{app:Ssym}), the 
asymptotic memory kernel (\ref{app:MemoryKernel})
becomes
 \begin{align}
M_{\alpha\al}(y,s) = 2 \left[(\dG^\al_{\mu \mu})^2 + (\dG^\al_{\mu, M-\mu})^2 + (\dG^\al_{M-\mu, \mu})^2 +(\dG^\al_{M-\mu,M-\mu})^2\right]
S_{\mu}(y,s)^2 ,  \quad \al>0.
\label{app:MemoryKernelUmbilic}
\end{align}
We thus arrive to the \textit{effective} mode-coupling equations for  $S_\al \equiv S_{\al \al}$,
\begin{equation}
\begin{split}
\partial_t S_\al(x,t) ={}& \big(-v_\al \partial_x + D_\al \partial_x^2\big)\, S_\al(x,t) \\
&+ \int_0^t \rmd s \int_{\R} \rmd y \; S_\al(x-y,\, t-s)\, \partial_y^2 M_{\al\al}(y,s),
\end{split}
\qquad \al = 0, 1, \ldots, \tfrac{M-1}{2}.
\label{modecoupling}
\end{equation}
with the modified memory kernel for umbilic modes given by (\ref{app:MemoryKernelUmbilic}).
Note that the range of $\al$ now extends from $0$ to $(M-1)/2$ instead of the initial range $0,\ldots ,M-1$.
The system of equations \eqref{modecoupling} 
with the definitions (\ref{app:MemoryKernelUmbilic}) constitute a renormalized set of 
mode-coupling equations,  adapted for
doubly degenerate modes.  
We can then analyze  Eqs.~\eqref{modecoupling} in a usual way using (\ref{app:DynExponents}),  which yields the dynamical exponents and scaling functions for all conserved modes.
 In particular, for our minimal model depicted in Fig.~\ref{fig:propagation}, we get the effective memory kernels 
\begin{align*}
&M_{00}(y,s) = \rho(1-\rho) (1-2\rho)^2 ((S_1(y,s))^2 + (S_2(y,s))^2),\\
&M_{11}(y,s) = \frac{\rho(1-\rho)}{2} (1-2\rho)^2(S_2(y,s))^2,\\
&M_{22}(y,s) = \frac{\rho(1-\rho)}{2} (1-2\rho)^2 (S_1(y,s))^2.
\end{align*}
From the effective memory kernels we predict GR universality for all modes,  with the dynamical exponent 
$z_0=z_1=z_2 = \varphi$, and scaling function given by Eq.~\eqref{Scalingfunction}. 
The respective asymmetricity parameters $\be$ for  the scaling function (\ref{Scalingfunction}) for different modes $\al$ follow by taking into account the modes velocities \cite{Popkov2016}.

\end{document}